\documentclass[aps,prl,twocolumn,groupedaddress]{revtex4-1}

\newcommand{\etc}{\textit{etc.}}
 \newcommand{\ie}{\textit{i.e.\ }}
 \newcommand{\eg}{\textit{e.g.\ }}

 \newcommand{\Eq}[1]{Eq.\ (\ref{#1})}

\begin{document}


\title{Universal scaling of resolution with photon number in superresolution fluorescence microscopy}


\author{Alex Small}
\email[]{arsmall@csupomona.edu} \homepage[]{\\
sites.google.com/site/physicistatlarge}
\affiliation{Department of Physics and Astronomy, California State
Polytechnic University, Pomona, CA (USA)}


\date{\today}

\begin{abstract}
Superresolution fluorescence microscopy techniques beat the
diffraction limit, enabling ultra-high resolution imaging in
biological physics and nanoscience.  In all cases that have been
studied experimentally, the resolution scales inversely with the
square root of some parameter that measures the number of photons
used. However, this ubiquitous limit arises from very distinct
mechanisms in different approaches, raising the question of whether
it is a fundamental limit that cannot be exceeded, or merely a
coincidence of the techniques studied thus far.  We demonstrate
that, under very general assumptions that encompass essentially all
fluorescence microscopy situations, the known resolution limit is
indeed universal.  Our model considers experiments that build up an
image via any arbitrary sequence of steps compatible with our
assumptions of (1) light that exhibits shot noise and (2) molecules
that can be modeled with rate equations.  A detailed examination of
our assumptions shows that exceeding this resolution limit will
require the use of quantum optical effects, pointing to an avenue
for future innovation.
\end{abstract}


\maketitle

Superresolution techniques\cite{Dyba2002,Gustafsson2005, Lidke2005,
Betzig2006, Rust2006, Hess2006} can beat the diffraction
limit\cite{Abbe1873} in fluorescence microscopy, providing important
tools for biological physics \cite{Wang2008,Greenfield2009,Hess2007}
and nanoscience \cite{Brimhall2011, Xu2009}.  While superresolution
can arise from a variety of different mechanisms, the key difference
between superresolution and conventional fluorescence microscopy is
that fluorescent molecules are not simply illuminated and read out.
Instead, they are either controlled
deterministically\cite{Dyba2002,Gustafsson2005}, or else a
stochastic control scheme is accompanied by substantial
post-processing\cite{Lidke2005, Betzig2006, Rust2006, Hess2006}.  In
superresolution fluorescence methods the resolution scales as the
wavelength $\lambda$ divided by the square root of a parameter
proportional to the number of photons $N$ used in the
experiment\cite{Hemmer2012}. However, this ubiquitous
$\lambda/\sqrt{N}$ limit arises from different mechanisms in
different cases, raising the question of whether it is universal or
merely coincidental. Some theoretical work has considered how the
performance of stochastic methods is limited by several different
factors\cite{Fitzgerald2012, Small2009a}, but the universality of
the inverse square root scaling law remains an open question. Here
we show that this limit is indeed universal for any superresolution
fluorescence microscopy technique built from an arbitrary
combination of elementary steps if the experiment involves (1) light
exhibiting shot noise and (2) molecules whose states can be modeled
with simple rate equations.

With stochastic switching (PALM, STORM, \etc\
\cite{Lidke2005,Betzig2006, Rust2006, Hess2006}), an image is built
by estimating individual molecular positions in each frame. The
precision scales as $\lambda$ over the square root of the number of
photons collected\cite{Ober2004}: Conceptually, one is determining
the mean position of $N$ photons at the detector. These positions
are independent random variables with standard deviation $\propto
\lambda$, so the standard deviation of the mean scales as
$\lambda/\sqrt{N}$. (A more rigorous derivation of this result
invokes the proportionality between Fisher information and photon
emission rate when the light sources exhibit shot
noise.\cite{Ober2004}) In methods that rely on saturation of a
transition (\eg\ STED\cite{Hell2007}, SSIM\cite{Gustafsson2005}),
the parabolic profile near the node of the illumination beam results
in resolution scaling as $\lambda$ divided by the square root of an
illumination intensity:  Near the node, the intensity profile is a
quadratic function of displacement, and detectable changes in signal
occur over a distance given by $r^2 k^2 I_0 = I_{\text{sat}}$, where
$r$ is distance from the node, $k=2\pi/\lambda$ is the wavenumber,
$I_0$ is proportional to the illumination power (\ie\ the number of
photons hitting the sample in a given time) and $I_{\text{sat}}$ is
the intensity at which the population in some energy level
saturates. Consequently, the smallest resolvable feature scales as
$\lambda/\sqrt{I_0}$\cite{Harke2008, Schwartz2009}.

We consider fluorescent molecules whose states respond to excitation
beams in a manner describable with simple rate equations, and are
read out information by detecting light exhibiting shot noise. The
shot noise assumption excludes the use of $N$ entangled
photons\cite{Boto2000}, where resolution can scale as $\lambda/N$.
The rate equation assumption excludes detection of molecular
positions via Rabi oscillations\cite{Sun2011}. We do not explicitly
consider negative index materials\cite{Pendry2000} or superresolving
pupils\cite{Difrancia1952, Huang2009}.  However, our analysis
applies to these technologies if the width of the point spread
function (PSF) replaces $\lambda$: The focal spots have finite width
in any real implementation, and near minima and maxima the intensity
must be a quadratic or higher-order function of position, so that
the field has a continuous second derivative in the wave equation.
Our analysis here only requires focal spots of finite width, with
intensity profiles that are parabolic near minima and maxima.

We begin by considering generalizations of deterministic
superresolution methods, \eg\ STED and SSIM. In deterministic
methods, superresolution is achieved by saturating a transition and
reading out spontaneous emission from an excited state. Both STED
and SSIM require only a single absorption event and a single
downward radiative transition. We will consider whether it is
possible to get resolution scaling as $\lambda/I_0^m$, for some
power $m > 1/2$, by shuffling the molecules through a sequence of
many transitions before read-out of information via detection of
spontaneous emission.

We assume molecules with an arbitrary set of energy levels,
arbitrary lifetimes for radiative and non-radiative transitions, and
arbitrary absorption cross-sections.  Molecules are present in a 2D
sample at a density $n(x,y)$, and light is read out in discrete
steps by a scanning lens (assumed to be diffraction-limited) that is
focused at $\mathbf{r_0}=(x_0,y_0)$; the detector is at infinity to
collect light from the smallest possible region. 3D sample depth
will not be considered; the chief effect would be to contribute an
out-of-focus background, and our effort here is to produce a
best-case limit. We allow for the possibility of detection in
multiple spectral channels, to distinguish different transitions of
interest, and we allow for the possibility of time-resolved
detection to distinguish processes with different lifetimes.

In a given spectral channel $i$ (corresponding to spontaneous
emission from a given transition) at a time $t$, one detects the
signal $S_i$:
\begin{eqnarray}
    S_i(t) = \int_{\text{sample}} n_i(\mathbf{r'},t) h(\mathbf{r_0-r'}) d^2\mathbf{r'} =
    h*n_i
\end{eqnarray}
where $n_i$ is the density of molecules in excited level $i$, $h$ is
the PSF of the collection lens, and $*$ denotes convolution.  To
resolve a spatially inhomogeneous structure, one must look at
\textit{changes} in signal from one point to the next. The relevant
quantity is:
\begin{equation}
    \frac{\partial S_i}{\partial x} = h*\frac{\partial n_i}{\partial x}
    \label{eq:signal}
\end{equation}
We will therefore be most interested in the regions of the sample
where $n_i$ changes most rapidly.

We also assume that the molecules are illuminated by some arbitrary
set of beams, each with frequency $\omega_j = ck_j$ (where $k_j$ is
the wavenumber of the beam) chosen to be tuned to some transition of
the molecule.  The beams are focused at positions $(x_j,y_j)$, not
necessarily coinciding with the focus of the detection lens at
$(x_0,y_0)$, and have intensity profiles of the form:
\begin{eqnarray}
    I_j (x,y,t) = I_0 a_j(t) f_j(x-x_j,y-y_j)
    \label{eq:intensities}
\end{eqnarray}
$f_j(x,y)$ is the square of some non-evanescent solution to the wave
equation.  $a_j(t)$ represents a possibly time-dependent modulation
of the intensity, \eg\ to perform STED by first raising molecules to
the excited state and subsequently sending most of them to the
ground state, or to switch a beam on and off to probe different
transitions at different times. Allowing modulation of beam
intensities means that we may be interested in integrals of $S_i(t)$
over specified time intervals. $I_0$ is an overall scaling
parameter; it enables us to take a high-intensity limit by tuning a
single parameter rather than treating each beam separately.
Crucially, $I_0$ is proportional to the number of photons incident
on the specimen.

We assume that the kinetics of the molecule can be modeled with rate
equations.  The temporal behavior of the level occupations $\{
n_i(x,y,t) \}$ will be exponentially-decaying transients plus a
steady-state:
\begin{eqnarray}
    n_i(x,y,t) = n_i^{(s)}(x,y) + \sum_{\text{transients } \beta} n_i^{(\beta)}(x,y) e^{-t/\tau_{\beta}}
    \label{eq:populations}
\end{eqnarray}
where $\beta$ indexes the transients and $n_i^{(s)}(x,y)$ is the
steady-state and $\tau_{\beta}$ is the lifetime of the transient
$\beta$. The spatial dependence of $n_i^{(s)}(x,y)$ and
$n_i^{(\beta)}(x,y)$ is determined by the local values of beam
intensities. Depending on how detection is time-gated, and how the
intensities are modulated via $\{a_i(t)\}$ in \Eq{eq:intensities},
our signal $S_i$ may be dominated by the local value of either
$n_i^{(\beta)}$ or $n_i^{(s)}$.

As we increase $I_0$, irrespective of whether we are detecting a
transient signal or a steady-state signal, the relevant coefficient
in \Eq{eq:populations} saturates at some limiting value that is
independent of $I_0$.  Consequently, $n_i$ can only depend on
\textit{ratios} of local beam intensities. These ratios vary on
length scales of $\approx \lambda$ everywhere except near the nodes
of beams. At a node, molecules do not ``see'' the beam, and very
close to the node the population is in a weak-field limit (with
respect to that beam) rather than an asymptotic strong-field limit.
Thus, the most rapid spatial variation of the coefficients in
\Eq{eq:populations} occurs near nodes, where there's a cross-over
between different limiting behaviors. The widths of cross-over
regions can be determined by assuming that near the nodes the
intensity is a quadratic function of position:
\begin{eqnarray}
    I_j(x,y,t) \approx I_0 a_j(t)k_j^2 \left( c_x x^2+ c_y y^2 \right)
\end{eqnarray}
(We assume a coordinate system in which the quadratic form is
diagonalized.)  The cross-over happens when the intensity is
comparable to some saturation intensity $I_{\text{sat}}$.  For
displacements away from the node in the $x$ direction the cross-over
happens at:
\begin{eqnarray}
    \delta x \approx \frac{1}{k_j}\sqrt{\frac{I_{\text{sat}}}{I_0 a_j c_x}}
    = \frac{\lambda_j}{2\pi}\sqrt{\frac{I_{\text{sat}}}{I_0 a_j
    c_x}} \propto \frac{\lambda_j}{\sqrt{I_0}}
\end{eqnarray}
We thus get that the length scale over which the signal changes
rapidly, and hence the length scale of the features in the data, is
proportional to $\lambda$ divided by the square root of a measure of
the number of photons used.

If the beam profile is non-parabolic near the node (\eg\ $r^4$) we
could proceed similarly, but instead of getting $1/\sqrt{I_0}$ in
our result we would get $1/I_0^{1/4}$ or some other (lower) power of
$I_0$. This width would decrease more slowly for $I_0\rightarrow
\infty$, giving worse scaling between resolution and intensity. One
cannot use a node where the intensity scales as $x^{n}$ ($n<2$), as
that would imply an electric field that scales as $x$ to a power
$<1$, giving a discontinuous derivative in the wave equation.

In considering whether a structure can be resolved, we must also ask
whether translating the collection lens by a distance $\delta x$
produces a change in signal greater than the fluctuations of the
noise in the signal. We get a condition for the smallest resolvable
feature if we equate the change in signal $\delta x \frac{dS_i}{dx}$
with the square root of the signal (assuming shot noise).  For large
$I_0$, the spontaneous emission rate saturates at one photon per
excited state lifetime $\tau$, so the signal saturates at a value
proportional to $\Delta t/\tau$, where $\Delta t$ is the acquisition
time.  The derivative of the signal scales as $S_i\sqrt{I_0}/\lambda
$. Putting this together gives:
\begin{eqnarray}
    \delta x \propto \frac{\lambda}{\sqrt{I_0 \Delta t}}
    \label{eq:spontlimit}
\end{eqnarray}
We thus see that the smallest resolvable feature size scales
inversely with the square root of a measure of the number
illumination photons ($I_0$) and also the square root of a measure
of the number of photons collected in the experiment ($\Delta t$).

According to \Eq{eq:spontlimit}, if we examine the Fourier transform
of an image built by scanning and collecting spontaneous emission in
our scheme, the largest spatial frequency component distinguishable
from noise is $k_{\text{max}} \propto (\sqrt{I_0 \Delta
t})/\lambda$.  If we were to try to extract additional information
by taking linear combinations of measurements at different
positions, the Fourier transform of the image built from these
linear combinations will still have a finite width in frequency
space, scaling as $(\sqrt{I_0 \Delta t})/\lambda$. Alternately, one
could take nonlinear combinations of signals, \eg\ multiply signals
shifted in time or space\cite{Dertinger2009}.  In position space,
the key quantities of interest will be peaks in either the nonlinear
combination or a spatial derivative thereof. If peaks have quadratic
maxima, we can approximate them locally with Gaussians.  Multiplying
$m$ Gaussians gives a function of the form $e^{-mx^2/\sigma^2}$,
with width $\sigma/\sqrt{m}$. If we work with signals shifted in
time, the factor $m$ is to the number of times that a measurement is
performed, and is hence again proportional to the number of photons
used in the experiment.  We thus conclude that post-processing
cannot improve the scaling between resolution and photon count.

Instead of detecting spontaneous emission, one could also detect
photons emitted via a coherent response to the external driving
field, \eg\ spontaneous emission\cite{Gather2011} or nonlinear
processes like harmonic generation and CARS\cite{Cheng2006,
Xie2006}. Nonlinear microscopy is usually performed far from a
saturated regime, \ie\ in a regime in which the response of the
specimen can be modeled as either a power of the incident intensity
(in harmonic generation) or a product of different beam intensities
(\eg\ in CARS). For unsaturated nonlinear microscopy with a single
beam or multiple co-focused beams, the resolution is known to be
enhanced by only a factor of $1/\sqrt{m}$, where $m$ is the order of
the nonlinearity (number of simultaneously absorbed photons), due to
the parabolic nature of the intensity maxima\cite{Fukutake2010}.
However, a scaling of signal as $I_0^m$ ($m>1$) cannot be sustained
for arbitrarily large incident powers; eventually energy
conservation would be violated.

If coherence is maintained in the saturated regime, the detected
intensity is \textit{not} added linearly from the different regions
of the focal area.  Instead, the amplitude $A$ is a coherent sum of
contributions from different parts of the sample. The amplitude at
the surface of the detector can be described by an amplitude Point
Spread Function (aPSF\cite{Fukutake2010}). At the detector, the
local amplitude is the aPSF-weighted sum of the local fields at each
point in the focal region.  We can easily extend the treatment in
the previous section to cover this case, assuming again illumination
by some arbitrary set of beams, each having an amplitude
proportional to $\sqrt{I_0}$, and in the vicinity of a node the
amplitude of each field component is a linear function of the
displacement from the node.

Each point in the specimen contributes to the signal amplitude in an
amount $dA$, and in the limit of large $I_0$ energy conservation
requires that $dA$ is proportional to $\sqrt{I_0}$. The ratio of
$dA$ to $\sqrt{I_0}$ saturates as a function of local beam
amplitudes, varying rapidly only near nodes, as discussed above.  As
the beam is scanned, the largest change in signal thus occurs when a
node is scanned through the position of a molecule.  By the same
arguments as above, the largest changes in signal happen in a region
of size $\delta x \propto \lambda/\sqrt{I_0}$.

As before, we also need to consider whether the change in signal
exceeds the noise.  The signal intensity now saturates at a value
proportional to $I_0 \Delta t$ rather than (in the spontaneous
emission case) a value independent of $I_0$, so the noise is
proportional to $\sqrt{I_0} \Delta t$.  We thus set $\delta x I_0
\Delta t/(\lambda/\sqrt{I_0}) = \sqrt{I_0 \Delta t}$ and get:
\begin{eqnarray}
    \delta x \propto \frac{\lambda}{I_0 \sqrt{\Delta t}}
    \label{eq:coherent}
\end{eqnarray}
The denominator now contains a factor proportional to the number of
photons \textit{incident} in the experiment.  However, the
resolution still scales inversely with the square root of the number
of photons \textit{detected}.  This is the key difference between
the cases of spontaneous and stimulated transitions:  Because the
photon emission rate is no longer bounded by the inverse lifetime of
a state, a larger number of photons can be collected in a time
$\Delta t$.

Let us now consider localization-based approaches, which typically
use one\cite{Folling2008} or two\cite{Betzig2006,Lidke2005,Rust2006,
Hess2006} illumination beams to perform the tasks of switching
molecules between activated and dark states and exciting
fluorescence from those molecules currently in the activated states.
Neglecting pixellation and out-of-focus background, the fundamental
limit to localization precision scales as $\lambda/\sqrt{N}$. There
are two ways that one might try to surpass this limit: One could
attempt to confine activation and excitation to a sub-$\lambda$
region (via some control scheme analogous to those considered
above), and use that confined activation as prior information on the
molecule's position, obtaining a maximum a posteriori
estimate\cite{Kay1993} of position. Alternately, one might attempt
to use a sequence of beams in a control scheme that increases the
product of the photon emission rate and the time spent in the
activated state before returning to the dark state.  In the later
case, the resolvable feature size will still be inversely
proportional to the square root of the number of photons collected,
but one can ask whether it would at least improve by increasing the
illumination intensity $I_0$.

In the first approach, using some sequence of illumination steps to
confine activation to a small region, the linear dimension of that
region will (as discussed above) scale as $\lambda/\sqrt{I_0}$. We
can approximate the prior information on the molecule position as a
function with a quadratic maximum with width $\propto
\lambda/\sqrt{I_0}$. The conditional likelihood of the data given
that the molecule is at $x$ is also known to have a quadratic
maximum with width $\lambda/\sqrt{N}$, where $N$ is the number of
photons detected \cite{Ober2004}. When these are multiplied to get
the posterior probability of the position given the
data\cite{Kay1993}, we get another function with a quadratic
maximum, and the second order coefficient in the expansion is:
\begin{equation}
    (\gamma_1 N + \gamma_2 I_0)(x/\lambda)^2
\end{equation}
where the $\gamma$ coefficients contain all necessary factors of
$\pi$, saturation intensities, \etc\ The width scales inversely as
the square root of a linear combination of $N$ and $I_0$, and so we
again have a localization precision that scales as $\lambda$ divided
by the square root of some measure of the number of photons used.

In the second approach, we can try to increase the number of photons
collected by either increasing the photon emission rate or
decreasing the rate of passage from an activated state (one that can
fluoresce) to the dark state (one that cannot fluoresce). In the
best case (stimulated emission) the photon emission rate is
proportional to $I_0$.  If return to the dark state is via a
stimulated transition, then the ratio of photon emission rate to
rate of return would be independent of $I_0$. Consequently, to
achieve the best possible scaling of resolution with $I_0$, one
would need a molecule that returns to the dark state via a
spontaneous transition.  The rate of return to the dark state will
hence be proportional to the probability of being in a bottleneck
state. A bottleneck state will be one that can undergo a spontaneous
transition either to the dark state or to another state that
undergoes a sequence of transitions that always lead back to the
dark state. The only remaining question, in terms of optimizing the
scaling of resolution with illumination intensity, is whether the
probability of being in a bottleneck state can be driven to zero.

If a molecule emits many photons before returning to the dark state,
we can assume that at any particular time the probability $p_b$ of
being in a bottleneck state is steady.  (This statement is
conditional on the knowledge that the molecule is not yet in the
dark state.) It follows that the rate of transitions (upward or
downward) into that state will be equal to the rate of transitions
out of the state. This requires solving equations of the form $p_b
(k_{\text{spont}} + I_0 k_{\text{induced}}) = R_{\text{spont}} +
R_{\text{induced}}$ where the $k$ parameters are rate constants for
spontaneous and induced transitions, and the $R$ parameters are
rates of transitions into a bottleneck state, summed over all states
that can reach it. We get $p_b=(R_{\text{spont}} +
R_{\text{induced}})/(k_{\text{spont}} + I_0 k_{\text{induced}})$.
For large $I_0$, the limiting value of $p_b$ is is non-zero since
$R_{\text{induced}}$ is proportional to $I_0$. Thus, the rate of
return to the dark state cannot be driven to zero. The total number
of photons emitted by an activated fluorophore can therefore only
scale as $I_0$, and the resolution of the reconstructed image will
scale as $\lambda/\sqrt{I_0}$.

In conclusion, we have shown that in any fluorescence microscopy
experiment that satisfies a few simple assumptions (conditions that
are ubiquitous in fluorescence experiments in biology and
nanoscience), the best achievable resolution scales as the
wavelength of light divided by the square root of a measure of the
number of photons used in the experiment (aside from one borderline
case).  Any further innovation with common fluorescence tools cannot
lead to improved efficiency of superresolution. Our analysis does
not consider coherent quantum effects, which are known to enable
resolution scaling inversely with photon number. Thus, beating the
limit of $\lambda/\sqrt{N}$ will require collaboration between the
biomedical optics and quantum optics communities. The feasibility of
using coherent quantum affects to achieve resolution scaling better
than $1/N$ requires a separate analysis.

\begin{acknowledgments}
This work was supported by a Teacher-Scholar award from
 California State Polytechnic University.  A portion of the work was
 conducted at the Kavli Institute for Theoretical Physics at the
 University of California, Santa Barbara.  This research was supported in part by the National Science Foundation under Grant No. PHY11-25915.
\end{acknowledgments}


%

\end{document}